\documentclass[fleqn,12pt,twoside]{article}
\usepackage{espcrc1}
\usepackage{graphicx}

\newcommand{\ket}[1]{| #1 \rangle }
\newcommand{\bra}[1]{\langle #1 | }

\title{Real-time Skyrme TDHF dynamics of giant resonances}
\author{Takashi Nakatsukasa
\address[Tsukuba]{Institute of Physics and Center for Computational Sciences,
      University of Tsukuba, Tsukuba 305-8571, Japan}
and Kazuhiro Yabana\addressmark[Tsukuba]}

\begin{document}

\maketitle

\begin{abstract}
Nuclear dynamics of giant resonances are investigated with
the real-time Skyrme TDHF method.
The TDHF equation is explicitly linearized with respect to variation
of single-particle wave functions.
The time evolution of transition densities are calculated for
giant dipole resonances.
The time-dependent densities of protons and neutrons
suggest that the dynamics of giant dipole resonance in neutron-rich nuclei
are significantly different from that in stable nuclei with $N\approx Z$.
\end{abstract}

\section{Introduction}
\label{}

Atomic nuclei exhibit a variety of collective modes of excitation.
In particular, the giant resonances always have been of central interest
in nuclear structure and reaction studies.
The giant resonances correspond to the most fundamental oscillations
in nuclei, in the sense that they exhaust major portions of the
energy-weighted sum-rule values and that
the nucleus strongly absorbs energy from an external field, acting as a whole.
Indeed, the giant resonances can be qualitatively described in terms of 
semiclassical hydrodynamical models \cite{RS80}.

Among many kinds of giant resonances,
the isovector giant dipole resonance (GDR) is the most famous and
exhausts almost 100 \% of the sum-rule value.
It is well explained by ordinary hydrodynamical models.
This is rather exceptional because quantitative description of other modes
requires a treatment as a Fermi liquid \cite{RS80}.
There are two famous hydrodynamical models for GDR:
the Goldhaber-Teller (GT) model \cite{GT48} and
the Steinwedel-Jensen (SJ) model \cite{SJ50}.
The GT model predicts $A^{-1/6}$ dependence of the GDR
frequencies which arises from the concept that the restoring force is
proportional to the nuclear surface area.
In contrast, the SJ model relaxes the assumption of
incompressibility, which leads to $A^{-1/3}$ dependence of the GDR
frequencies.
Experimental data are best fitted by a combination of these two \cite{BF75}:
In light nuclei, the data seem to indicate the $A^{-1/6}$ law, while the
$A^{-1/3}$ dependence becomes increasingly dominant for increasing values
of $A$.

The hydrodynamical models have a close connection to
the time-dependent Hartree-Fock (TDHF) theory.
In the limit of $\hbar\rightarrow 0$, the TDHF equation goes over into
the Vlasov equation.
Therefore, without the collision term,
the TDHF should provide a microscopic description of an appropriate
hydrodynamical model.
Recently, we have proposed the real-time TDHF
method combined with
the absorbing-boundary condition (TDHF+ABC method)
for a linear response function \cite{NY01,NY05}.
This may be regarded as an extension of the continuum random-phase
approximation (RPA), made applicable to a deformed system.
In this paper, we show real-time dynamics of the TDHF for GDR
and discuss how their properties change from
stable ($N\approx Z$) to unstable nuclei ($N \gg Z$).
As is discussed in the following sections,
we take a small-amplitude limit of the TDHF.
Although this is equivalent to the RPA,
the time-dependent snap shots of the TDHF wave packet may provide
an intuitive dynamical picture of the GDR.
It should be also noted that the TDHF provides a proper description
for low-lying modes as well, for which
quantum effects are so important that
the semiclassical hydrodynamical models are not applicable.

\section{Linearized TDHF in real time}

The HF ground state is assumed to be a Slater determinant which consists of
$A$ single-particle orbitals,
$\Phi_0(x_1,\cdots,x_A)=\det\{\phi_i(x_j)\}_{i,j=1,\cdots,A}$ with
$x=(\vec{r},\sigma,\tau)$.
Each single-particle orbital is determined by
\begin{equation}
\label{HF-eq}
h[\phi,\phi^*] \phi_i(x) =
\epsilon_i \phi_i(x)
\quad\quad\mbox{ for } i=1,\cdots,A ,
\end{equation}
where $h[\phi,\phi^*]$ is the single-particle Hamiltonian which depends on
$\phi_i(x)$ ($i=1,\cdots,A$) self-consistently.
The TDHF equation is obtained by replacing, in Eq. (\ref{HF-eq}),
$\epsilon_i$ by the time derivative $i\hbar\partial/\partial t$, and
$\phi_i(x)$ by the time-dependent wave function $\psi_i(x,t)$.
The TDHF equation is now linearized with respect to
variation of each single-particle wave function
and a time-dependent external field $v(x,t)$.
Substituting
$\psi_i(x,t)=(\phi_i(x)+\delta\psi_i(x,t))e^{-i\epsilon_i t/\hbar}$
into the TDHF equation, we have
\begin{equation}
\label{LTDHF-eq}
i\hbar\frac{\partial}{\partial t} \delta\psi_i(x,t)
= \left( h[\phi,\phi^*] -\epsilon_i\right) \delta\psi_i(x,t)
+ \delta h(t) \phi_i(x)
+ v(x,t) \phi_i(x),
\end{equation}
where $\delta h(t)\equiv h[\psi,\psi^*] -h[\phi,\phi^*]$ in the first order of
$\delta\psi_i(x,t)$.
If we put $\delta h(t)=0$, Eq. (\ref{LTDHF-eq}) gives unperturbed
particle-hole excitations with a fixed single-particle potential in
$h[\phi,\phi^*]$.
$\delta h(t)$ is nothing but the residual interaction in
the language of the energy representation.
The second term in the r.h.s. of Eq. (\ref{LTDHF-eq})
contains a dynamical effect
which comes from variations of the self-consistent one-body potential.

Equation (\ref{LTDHF-eq}) is equivalent to the well-known RPA equation
in the energy representation.
In practice, however,
there are some differences, advantages and disadvantages in each method.
For instance, the uncertainty in energy, $\Delta E$,
is inversely proportional to
the period of the time propagation $T$; $\Delta E\sim \hbar/T$.
Therefore, when we want to distinguish states nearly degenerate,
we need to propagate the wave functions for a long period of time.
In this case, the energy representation may be a better choice.
On the other hand, when we are interested in a bulk structure of excited
states in a wide range of energy,
calculations using the time representation becomes more efficient
than those with the energy representation.
The time-dependent calculation should be suitable for giant resonances,
since their energies are rather high and spread over a wide range of energy.

The transition density in the time representation is defined by
the density variation from its ground-state value,
\begin{eqnarray}
\label{delta-rho}
\delta\rho(x,x';t) &=& \rho(x,x';t)-\rho_0(x,x') \nonumber\\
 &=& \sum_{i=1}^A \left\{ \phi_i^*(x) \delta\psi_i(x',t)
                   +\delta\psi_i^*(x,t)\phi_i(x') \right\} .
\end{eqnarray}
In this paper, we are mainly interested in the spin-independent
diagonal part of Eq. (\ref{delta-rho});
\begin{equation}
\label{delta-rho-diag}
\delta\rho_\tau(\vec{r};t) 
 = \sum_{i=1}^A \sum_{\sigma}\left\{ \phi_i^*(x) \delta\psi_i(x,t)
                   +\delta\psi_i^*(x,t)\phi_i(x) \right\} .
\end{equation}
The expectation value of an operator
$\hat{F}(\vec{r},\tau)$ can be expressed as
\begin{equation}
\label{Ft}
F(t) = F_0 + \delta F(t)
= F_0 +\sum_\tau\int d^3r \hat{F}(\vec{r},\tau) \delta \rho_\tau(\vec{r};t) ,
\end{equation}
where $F_0$ is the ground-state expectation value.

The external field $v(x,t)$ in Eq. (\ref{LTDHF-eq})
can be chosen according to the purpose of the calculation.
In order to calculate the strength distribution in a wide range of energy,
an instantaneous external field, $v(x,t)=v(x)\delta(t)$, is suitable,
because this excites the system to states in all energies.
In contrast, if we adopt an oscillating field with a fixed frequency $\omega$,
$v(x,t)=v(x) \cos(\omega t)$,
the system is excited to a specific state with $E_{\rm x}=\hbar\omega$.
In this way, we can investigate dynamical properties of the specific state
in the time-dependent manner.

To calculate the strength function of the operator $\hat{F}(\vec{r},\tau)$,
\begin{equation}
S(\hat{F};E) = \sum_{E'} \delta(E-E')
  \left|\bra{\Psi_{E'}} \hat{F} \ket{\Phi_0}\right|^2,
\end{equation}
we adopt the external field $v(x)$ proportional to $\hat{F}(\vec{r},\tau)$.
Then, $S(\hat{F};E)$ can be obtained as the Fourier transform of the
expectation value of Eq. (\ref{Ft}).

Before showing results, let us discuss
a numerical difficulty related to presence of zero modes in nuclei.
The ``zero mode'' means zero-energy modes of excitation associated with
the spontaneous symmetry breaking in the HF states,
such as translation and rotation.
These modes should correspond exactly to the zero energy if the numerical
calculation is perfect.
However, a small numerical error and approximation may give imaginary energies
to these modes.
Since the time evolution of wave functions carries all the information of
the excited states, the presence of these imaginary-energy modes leads
to a kind of numerical instability to
prevent performing a long period of the time propagation.
Thus, we need to remove components of the zero modes from the time-dependent
single-particle wave functions $\delta\psi_i(x,t)$.
The zero modes can be constructed by operating the symmetry operator $\hat{P}$
and its conjugate one $\hat{Q}$ to the ground state.
For the translational case,
$\hat{P}$ is the total momentum operator and $\hat{Q}$ is 
the center-of-mass coordinate.

\section{Giant dipole resonances in stable and neutron-rich nuclei}

We now apply the method to GDR in even-even Be isotopes.
We calculated $B(E1)$ distribution for Be isotopes in Ref.~\cite{NY05}
using the SIII parameter set.
We found that the large deformation splitting in $^8$Be and $^{14}$Be,
because of the large quadrupole (prolate) deformation in the ground state
($\beta_2\approx 0.8$).
However, the width in $^{14}$Be is much larger than that in $^8$Be.
These results are robust and do not depend on the choice of the
Skyrme parameter set.
We also found that there is a significant low-energy $E1$ strength
around $E_{\rm x}=5$ MeV for $^{14}$Be \cite{NY05}.
However, the $E1$ strength and its peak position are rather sensitive to
the choice of the parameters.
Thus, in this section, we show time-dependent transition densities
for the main peak of GDR.

\begin{figure}[t]
\centerline{
\includegraphics[height=0.4\textheight]{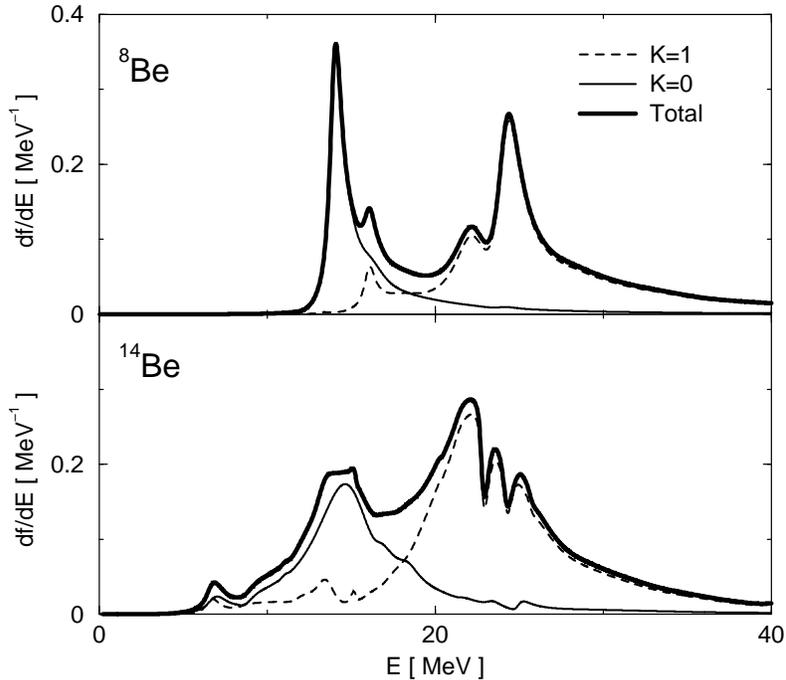}
}
\caption{Calculated $E1$ oscillator strength distribution in $^{8,14}$Be.
Thin solid and dashed lines indicate the response to dipole fields
parallel and perpendicular to the symmetry axis, respectively.
Thick line shows the total strength.
}
\label{fig:Be8_14}
\end{figure}

The calculation is performed on the three-dimensional Cartesian coordinate
grid space, using the Skyrme energy functional of Ref.~\cite{BFH87}.
The Galilean symmetry is respected in this functional including
the spin-orbit, Coulomb, and time-odd densities.
We adopt the SGII parameter set in the calculation.
In order to take account of the single-particle continuum, we use the
absorbing potential outside of the interacting region \cite{NY01,NY05}.
In Fig.~\ref{fig:Be8_14},
we show the $E1$ oscillator strength distribution for $^{8,14}$Be.
Two-peak structure due to the deformation splitting is prominent for
both $^8$Be and $^{14}$Be.
Hereafter,
let us focus our discussion on the peak around $E_{\rm x}=15$ MeV
with $K=0$.

\begin{figure}[t]
(p1)\hspace{90pt}(p2)\hspace{90pt}(p3)\hspace{90pt}(p4)\\
\includegraphics[height=0.24\textwidth]{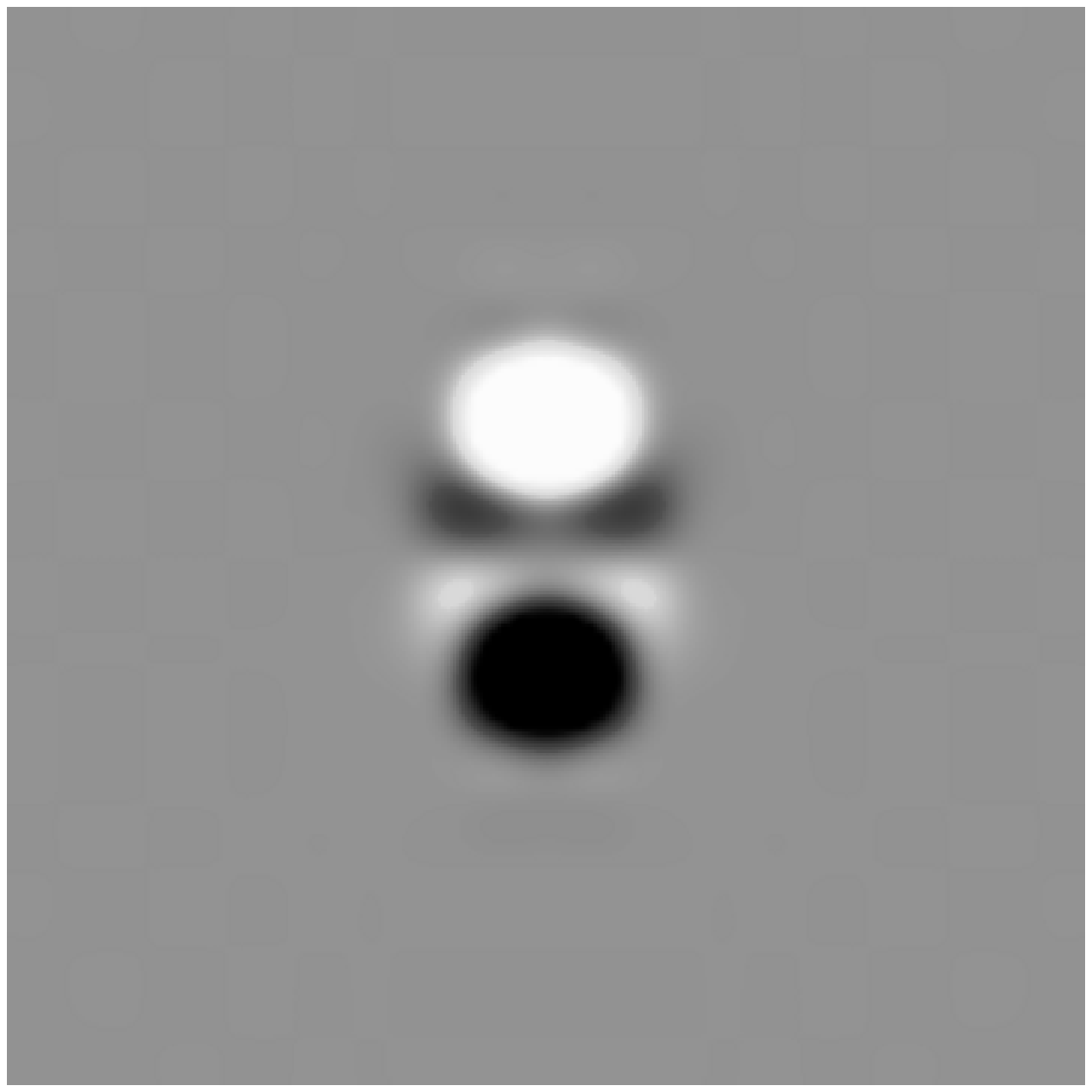}
\includegraphics[height=0.24\textwidth]{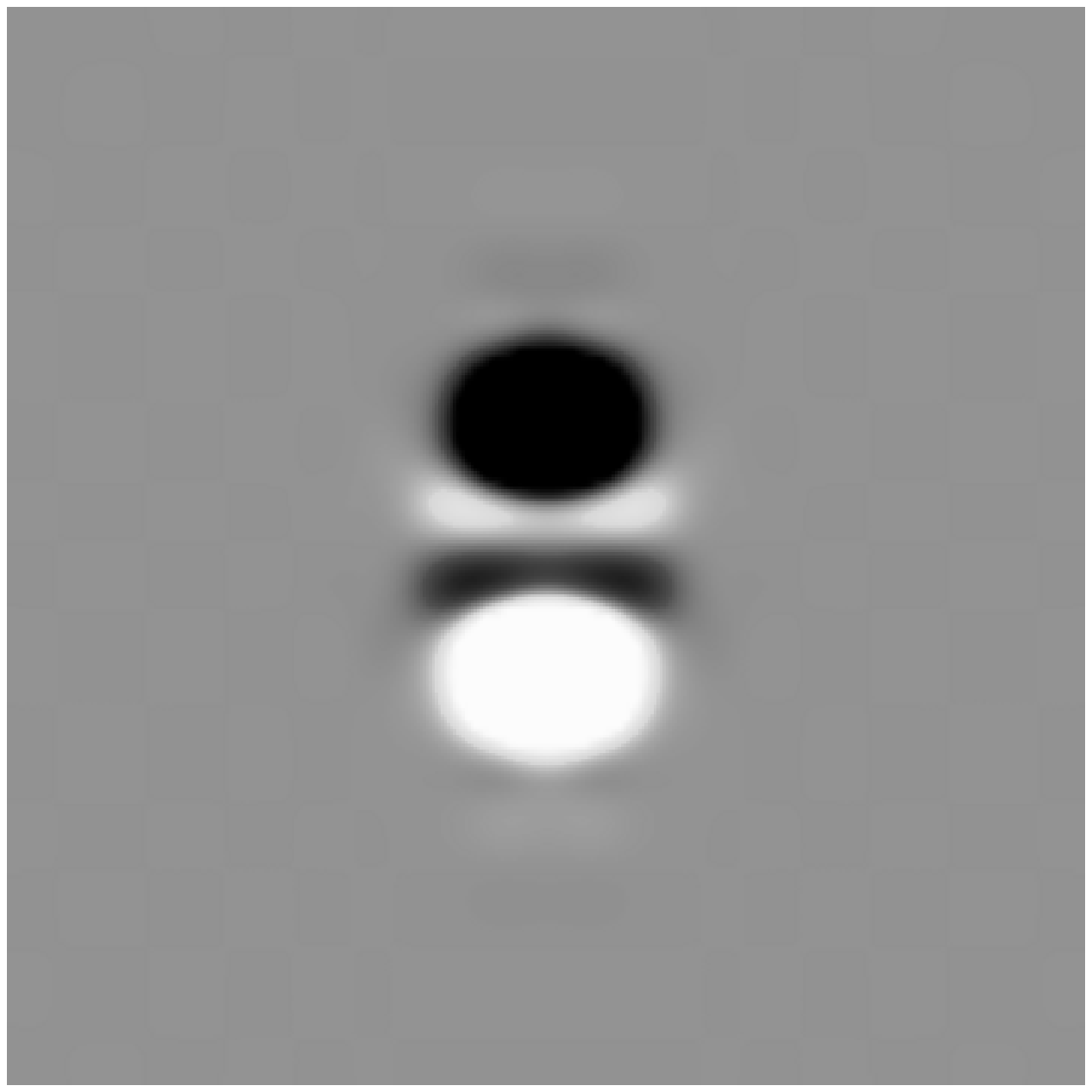}
\includegraphics[height=0.24\textwidth]{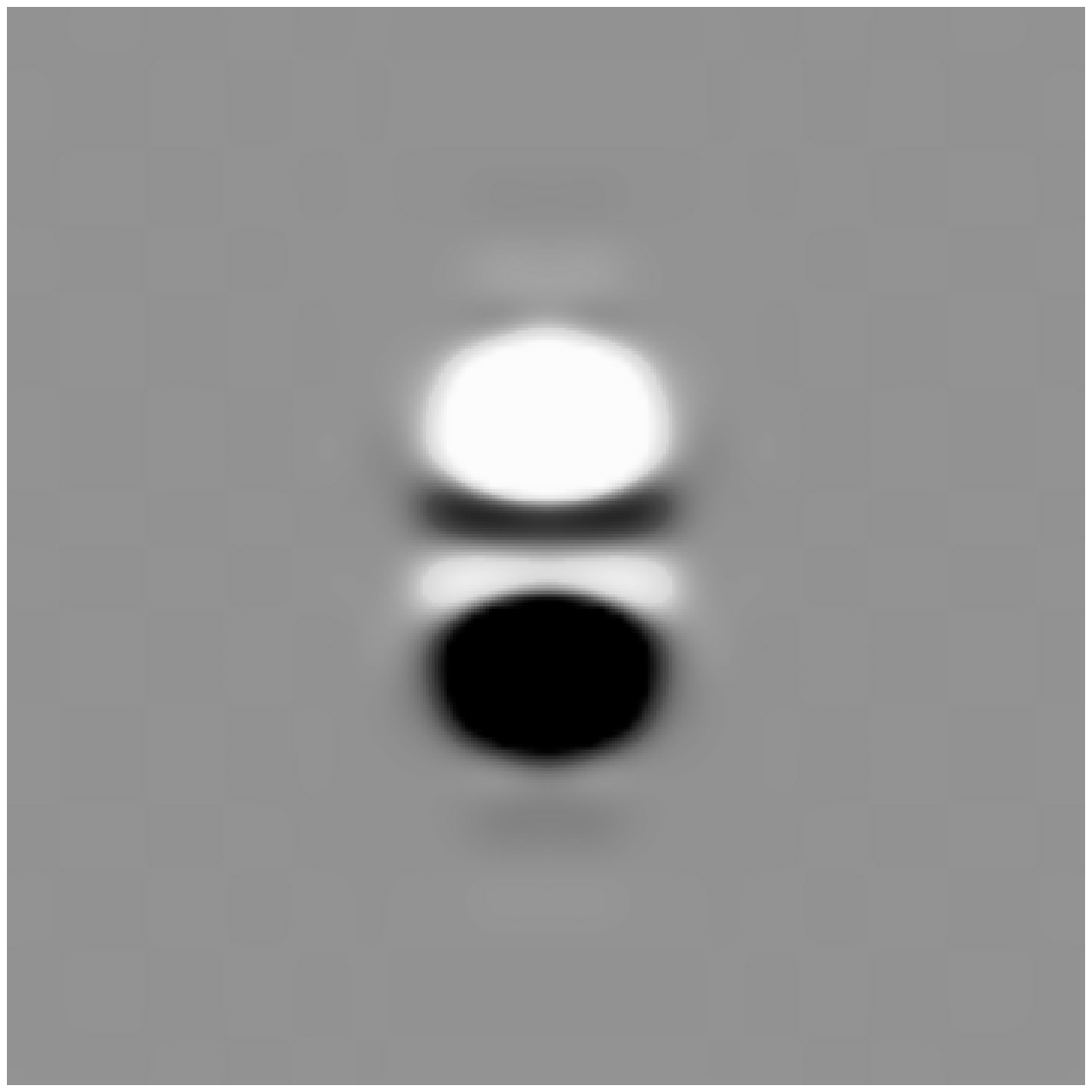}
\includegraphics[height=0.24\textwidth]{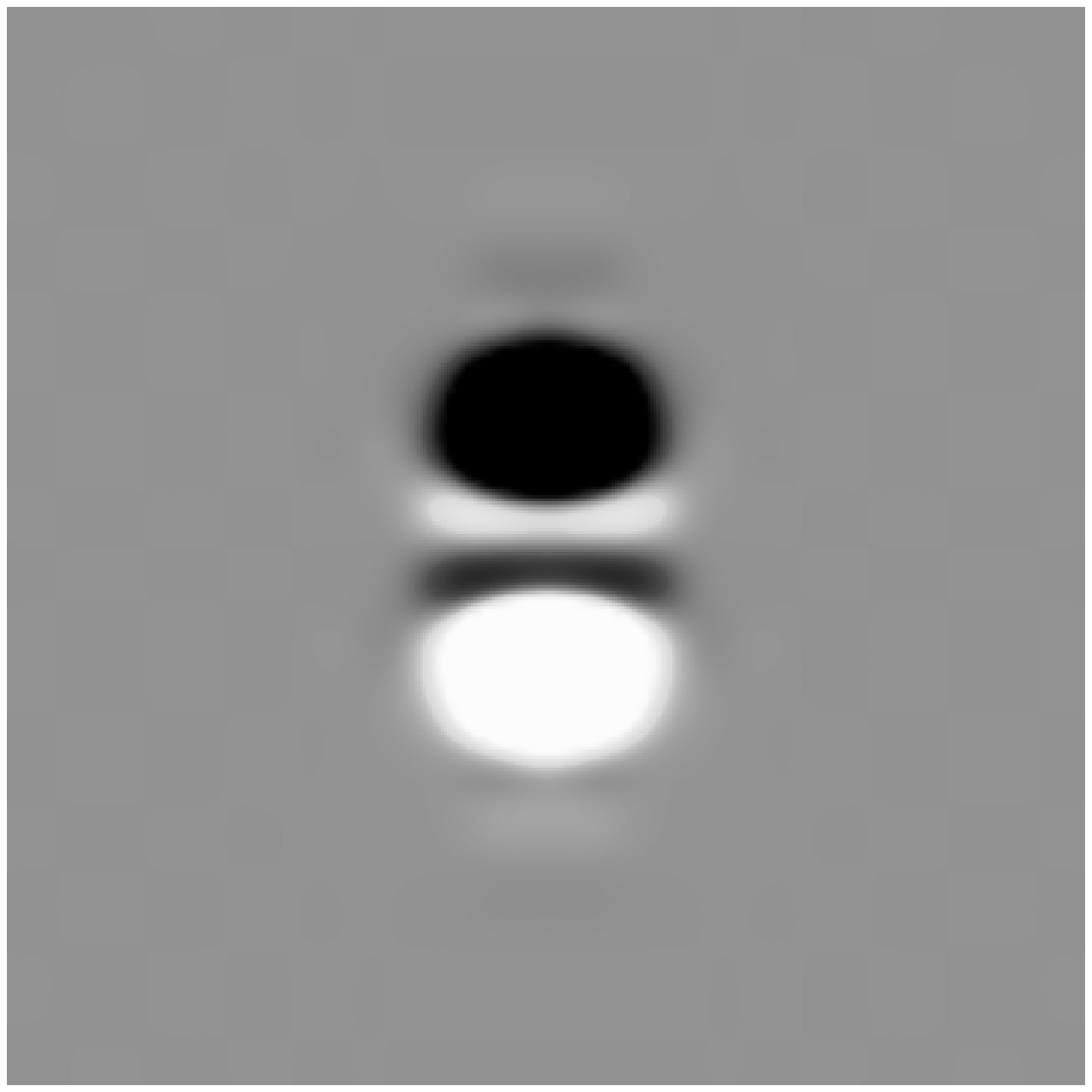}

(n1)\hspace{90pt}(n2)\hspace{90pt}(n3)\hspace{90pt}(n4)\\
\includegraphics[height=0.24\textwidth]{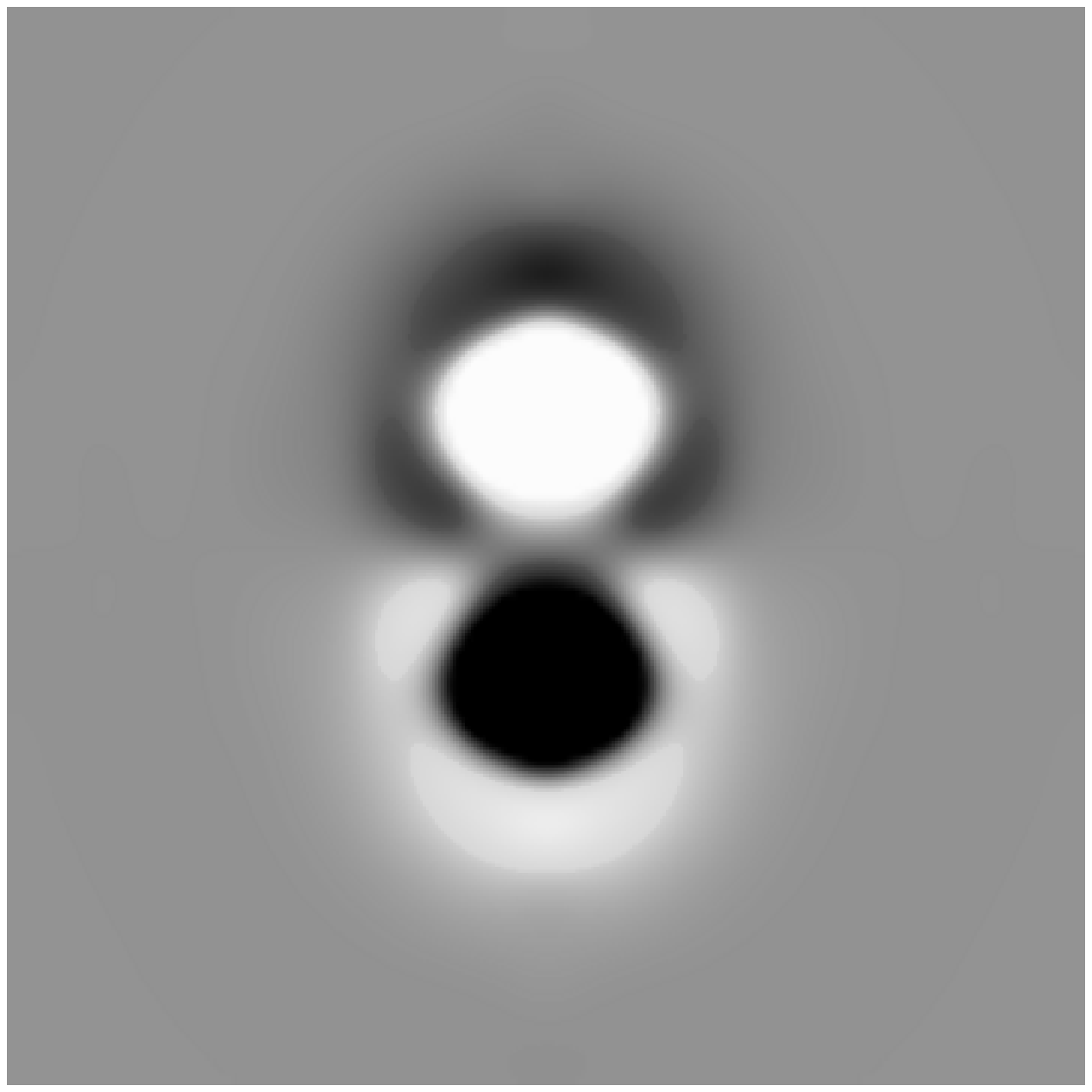}
\includegraphics[height=0.24\textwidth]{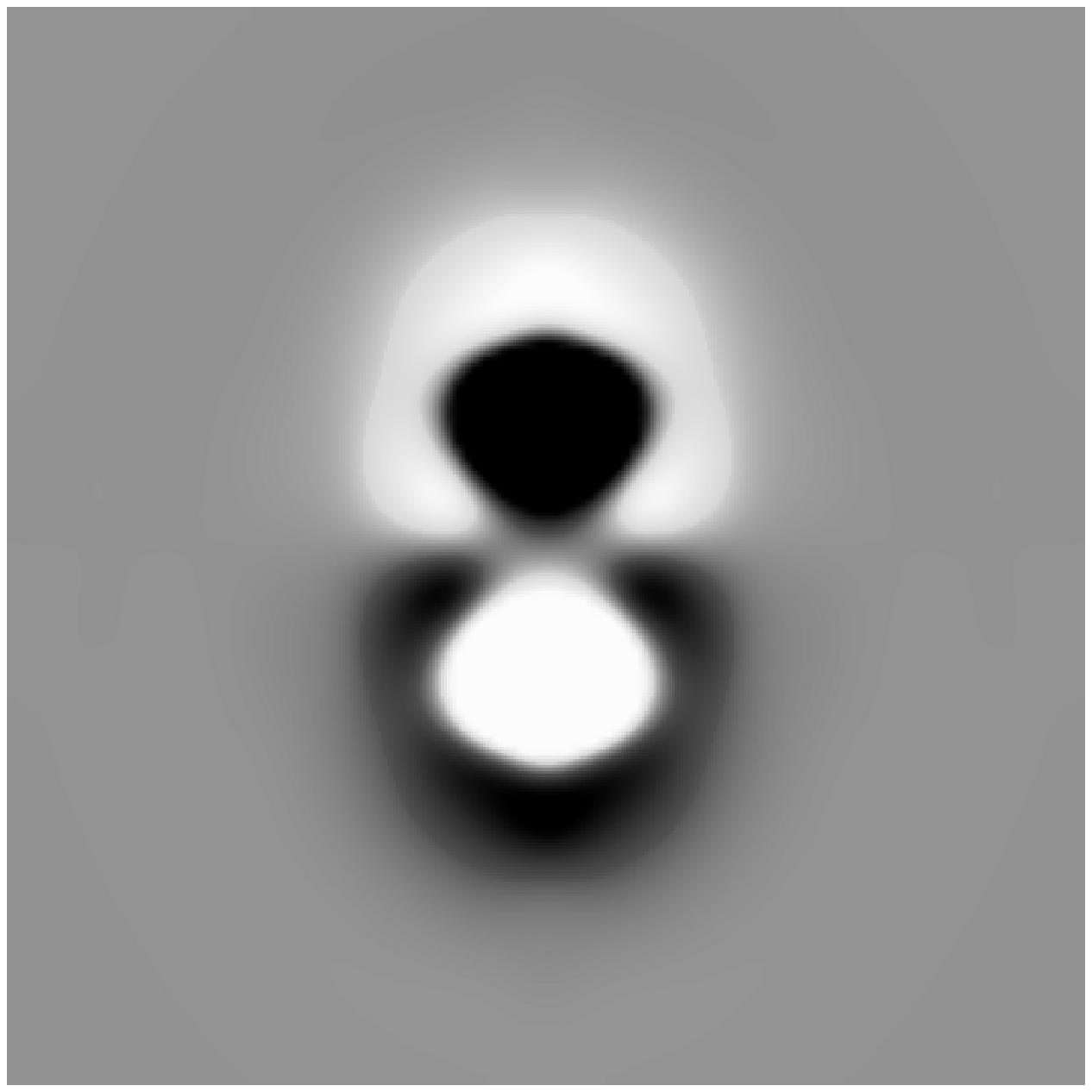}
\includegraphics[height=0.24\textwidth]{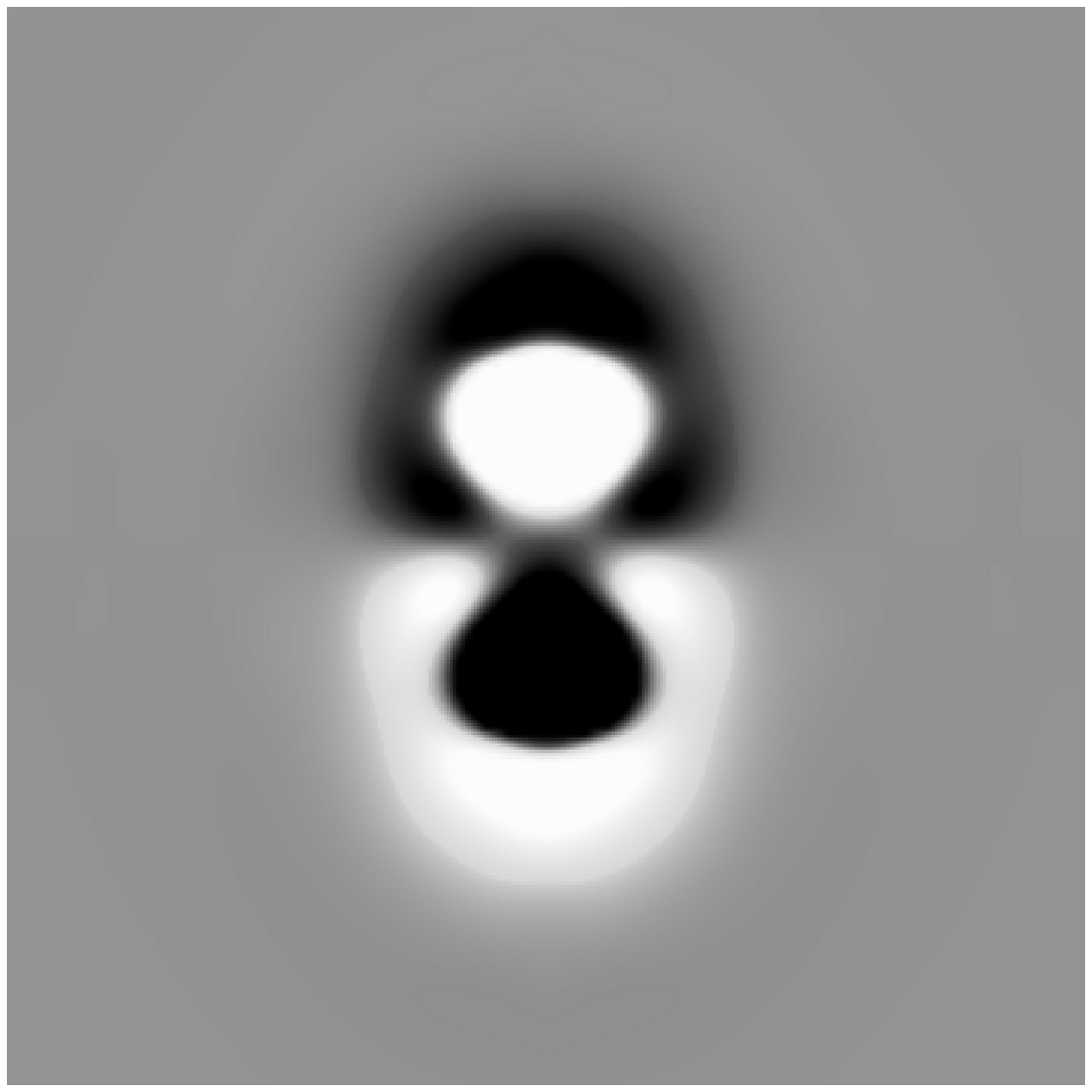}
\includegraphics[height=0.24\textwidth]{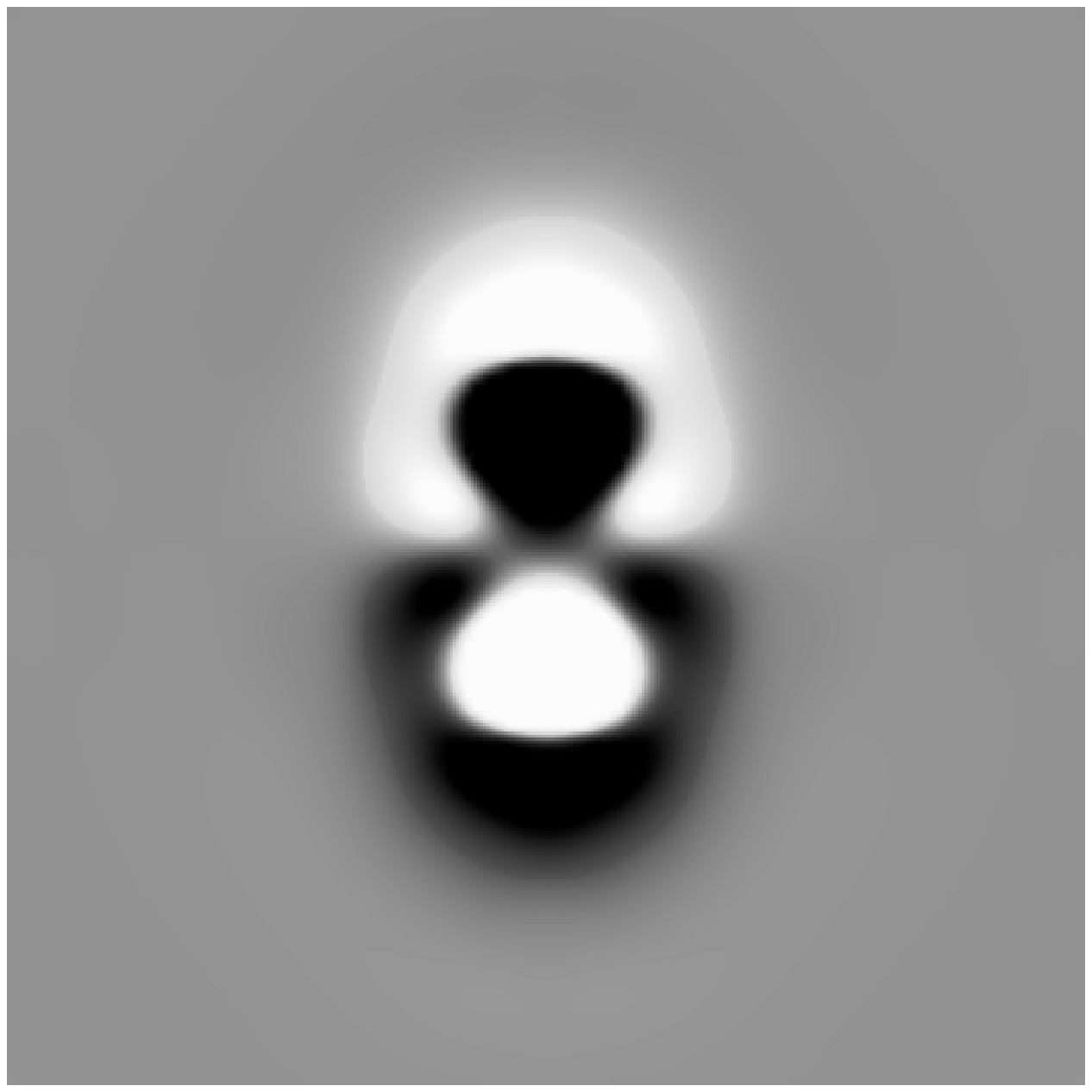}
\caption{Snap shots of calculated $\delta\rho_\tau(\vec{r},t)$
in the $x$-$z$ plane
for the $K=0$ peak at $E_{\rm x}=15$ MeV in $^{14}$Be.
The upper panels (p1-4) indicate $\delta\rho_p(\vec{r};t)$,
while the lower (n1-4) for $\delta\rho_n(\vec{r};t)$.
White (black) regions indicate those of $\delta\rho_\tau > 0$
($\delta\rho_\tau < 0$).
The time difference between two neighboring panels is
$\Delta t=0.2\ \hbar/$MeV.
The two panels at the same column corresponds to the same time $t$.
}
\label{fig:Be14}
\end{figure}

We use a Gaussian-pulse external field,
$v(x,t)=M(E1)_{K=0} \cos(\omega t) e^{-\gamma (t-t_0)^2}$, to
selectively excite the GDR around $E_{\rm x}=\hbar\omega=15$ MeV,
with $\gamma=3$ MeV$/\hbar$ and $t_0=2\ \hbar/$MeV.
Then, the spin-independent transition density of Eq. (\ref{delta-rho-diag})
is calculated in the 3D coordinate space.
It turns out that one of the Steinwedel-Jensen's assumptions,
$\delta\rho_n(\vec{r};t)=-\delta\rho_p(\vec{r};t)$,
is approximately satisfied for $^8$Be.
In contrast, in $^{14}$Be, we see a large deviation from this property.
Figure~\ref{fig:Be14} shows how $\delta\rho_\tau(\vec{r},t)$
($\tau=p,n$) evolve in time in the $x$-$z$ plane.
The time difference from one panel to the next
is $\Delta t = 0.2\ \hbar/$MeV which roughly corresponds to
the half period $\pi/\omega$.
We see that significant portions of neutrons actually move together with
protons.
The neutron transition density $\delta\rho_n$ shows a peculiar node structure.
In Fig.~\ref{fig:Be14}, the regions of $\delta\rho_p>0$ ($\delta\rho_p<0$)
have a large overlap with the those of $\delta\rho_n>0$ ($\delta\rho_n<0$).
This means a violation of the property of the SJ model,
$\delta\rho_p+\delta\rho_n=0$.
Weakly bound neutrons in neutron-rich nuclei seem to be significantly affected
by strong attraction between protons and neutrons,
and to oscillate in phase with protons' movement.
This is a consequence of the dynamical effect of
the time-dependent self-consistent potential,
$\delta h(t)$ in Eq. (\ref{LTDHF-eq}).

\section{Summary}

We present calculations of the linearized TDHF method in real time
for giant dipole resonances in $^{8,14}$Be.
These nuclei are calculated to be largely deformed in the ground state.
The main GDR peak is split into two peaks with $K=0$ and $K=1$.
We show the time-dependent transition densities for $K=0$ peaks.
The total density, $\rho_p+\rho_n$, is approximately conserved for $^8$Be,
while its conservation is significantly violated in $^{14}$Be.
The time evolution of the transition density, $\delta\rho_\tau(t)$,
suggests a strong dynamical effect for neutron-rich nuclei, and
seems to indicate the mixture of the isoscalar volume-type
and the isovector surface-type components.

This work has been supported by the Grant-in-Aid for Scientific
Research in Japan (Nos. 17540231 and 17740160).
The numerical calculations have been performed
at SIPC, University of Tsukuba,
at RCNP, Osaka University,
and at YITP, Kyoto University.

\end{document}